\documentclass{emulateapj}
\usepackage{graphics}
\usepackage{natbib}
\bibpunct{(}{)}{;}{a}{}{,}
\usepackage{amssymb}

\begin{document}

\title{Concentration of toroidal magnetic field in the solar tachocline by
$\eta$-quenching}

\author{Peter A. Gilman and Matthias Rempel}

\affil{High Altitude Observatory,
       National Center for Atmospheric Research\footnote{The National
       Center for Atmospheric Research is sponsored by the National
       Science Foundation} , 
       P.O. Box 3000, Boulder, Colorado 80307, USA
      }

\email{gilman@hao.ucar.edu,rempel@hao.ucar.edu}

\shorttitle{Concentration of toroidal magnetic field}
\shortauthors{Gilman \& Rempel}

\begin{abstract}
We show that if the turbulent magnetic diffusivity used in solar dynamos is 
assumed to be 'quenched' by increasing toroidal fields, much larger amplitude 
and more concentrated toroidal fields can be induced by differential rotation 
from an assumed poloidal field than if there is no quenching. This 
amplification 
and concentration mechanism is weakened and bounded by {\boldmath$j\times B$}
feedbacks on the 
differential rotation. Nevertheless, it is strong enough to contribute to the 
creation of $\sim 100$ kG toroidal fields near the base of the convection 
zone, perhaps 
in conjunction with the 'exploding flux tube' process. Such high fields are 
necessary for sunspots to occur in low solar latitudes. 
\end{abstract}


\section{Introduction}
There are substantial theoretical reasons for concluding that sunspots and 
active regions arise when toroidal flux tubes in or near the solar tachocline 
erupt to the photosphere \citep{Fan:etal:1994,Schuessler:etal:1994,
Caligari:etal:1995,Caligari:etal:1998}. 
Given the Sun's rotation rate, for these 
eruptions to occur in low solar latitudes where spots are found requires that 
the tubes in the tachocline have field strength approaching $100$ kG
\citep{Choudhuri:Gilman:1987}. Such large fields are a potential problem for 
dynamo theory applied to the Sun, since 
they are locally $10-100$ times the equipartition value when compared to the
kinetic energy density of both convective turbulence and differential 
rotation at tachocline depths. 

In all interface and flux transport dynamos applied to the sun, the strong 
toroidal fields achieved at tachocline depths come from the shearing by 
differential rotation there of a much weaker poloidal field. These toroidal 
fields are of much higher amplitude there than in the convection zone above 
because the magnetic diffusivity is assumed to be much smaller than in the 
much more turbulent convection zone. Since this shearing is also likely to 
lead to a back reaction on the local differential rotation, thereby 
extracting energy from it to amplify the magnetic field, the peak field 
achieved by induction depends critically on the rate energy can be resupplied 
to the differential rotation by the convection zone above. To be effective, 
this energy must be resupplied in a time quite short compared to a sunspot 
cycle. Since helioseismic inversions for 
differential rotation in the neighborhood of the tachocline do not show 
significant changes in the differential rotation within a solar cycle, 
whatever 
resupply is needed must be taking place. Indeed, the relative constancy of the 
observed differential rotation supports the validity of kinematic type dynamo 
models for the solar cycle, but sets constraints on the efficiency of the 
resupply mechanism and/or the toroidal field amplitudes produced by the shear. 
The resupply mechanism is not understood in detail, and any additional 
mechanism 
that helps produce concentrated high amplitude toroidal fields potentially 
reduces the amount of resupply needed and may allow higher toroidal fields 
to be produced for the same feedback.

A promising mechanism for achieving super-equipartition field strengths 
that does not depend totally on differential rotation is that of rising 
flux tubes 'exploding' before they reach the photosphere 
\citep{Moreno-Insertis:etal:1995,Rempel:Schuessler:2001}. In this 
scenario, many explosions of tubes result in significant loss of mass from 
inside the tubes, so that the part of the toroidal field that remains in 
the tachocline becomes more concentrated. If this happens at the unseen 
beginning of a new sunspot cycle, then the concentration of flux can lead 
to sufficient field strength that subsequent tubes get to the photosphere 
before they explode, so they can emerge as active regions.

Here we propose another, perhaps complementary, mechanism for concentrating 
toroidal magnetic flux, involving the effect of turbulence, and therefore 
turbulent diffusivity, being partially or even largely suppressed where the 
toroidal fields are strong. The effects of this "$\eta$-quenching" have been 
considered more broadly in dynamo theory previously by, for example, 
\citet{Ruediger:etal:1994} and \citet{Tobias:1996}, but 
without our emphasis on locally strong amplification of toroidal fields.
\citet{Ruediger:Kitchatinov:2000} included the effect of "$\eta$-quenching"
together with quenching of turbulent heat conductivity in a model for
the decay of sunspots. We use a formulation 
quite similar to that commonly used in dynamo models for 
"$\alpha$-quenching" \citep{Dikpati:Charbonneau:1999} (We make no attempt to 
evaluate the relative merits of various formulations of the $\eta$-quenching 
with guidance from MHD turbulence theory, such as is done in 
\citet{Vainshtein:Cattaneo:1992}). Our results below show 
that this $\eta$-quenching can be an extremely 
powerful mechanism for concentration when one is solving only the induction 
equation, but can be damped considerably when feedbacks on the differential 
rotation are taken into account. Thus this mechanism is subject to similar 
energetic limitations as when no $\eta$-quenching is allowed.

\section{Simple Model}
There are several levels of complexity that could be used to illustrate how
$\eta$-quenching works. We choose here to work primarily in Cartesian 
geometry, since that allows us to obtain exact analytical solutions to 
illustrate the effect. We have also used a Cartesian finite difference code 
to compute the time dependent solutions of the induction equation and the 
reaction of the induced {\boldmath$j\times B$} force on the 
differential rotation.

\subsection{Governing Equations}
\begin{figure}
  \resizebox{\hsize}{!}{\includegraphics{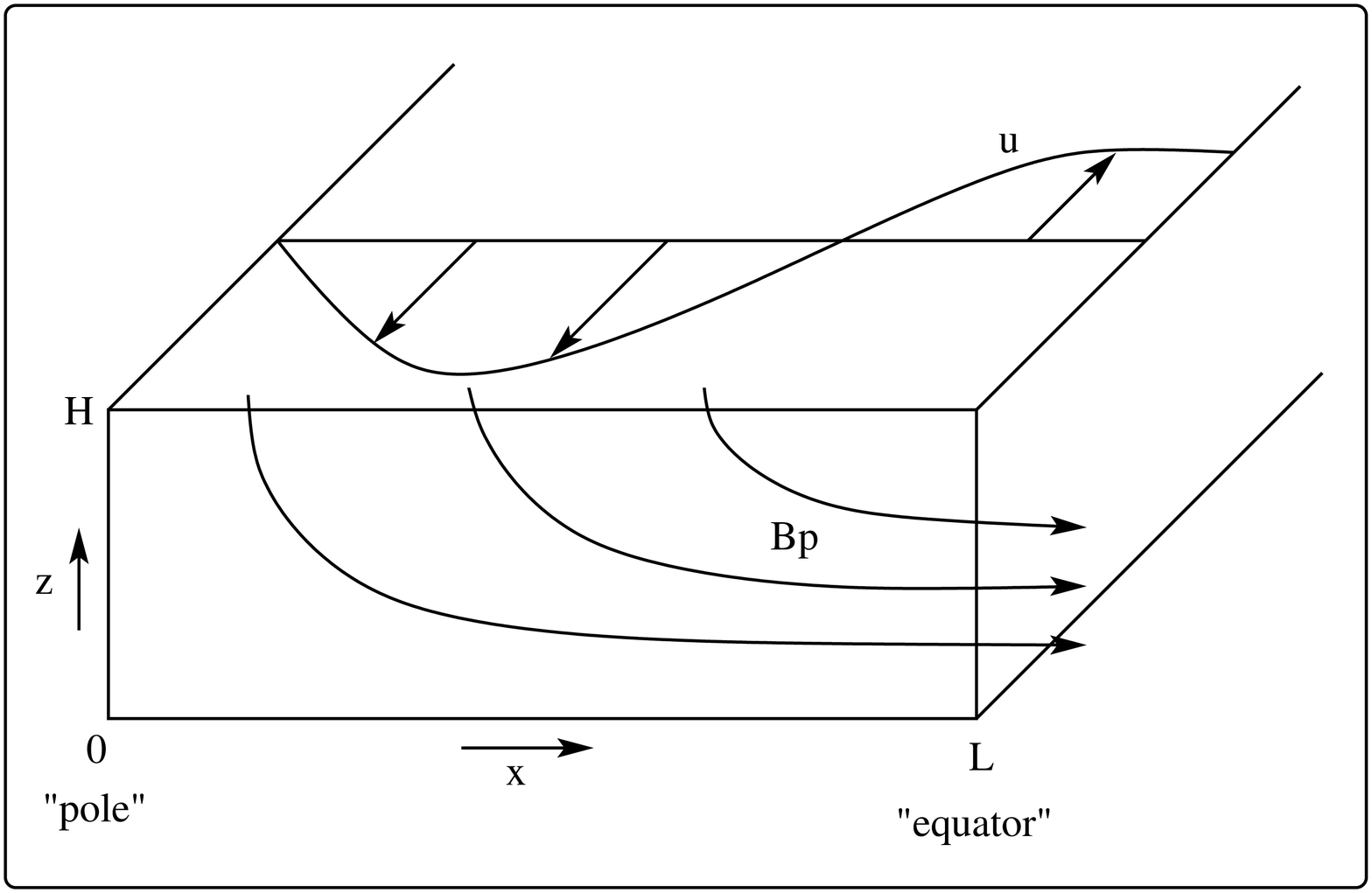}}
  \caption{Schematic view of the geometry of our domain. Indicated are also
    the imposed the velocity field $u$ and the poloidal magnetic field $B_p$.
  }
  \label{f1}
\end{figure}
For our model, we consider a channel that has an $x-z$ cross section, 
infinite in the down channel coordinate. Sidewalls are at $x=0$, $L$, 
bottom and top at $z=0$, $H$.  Roughly speaking, the $x$ coordinate 
corresponds to colatitude on the Sun, the $y$ coordinate, longitude and the 
$z$ coordinate the vertical. The "pole" would be at $x=0$, the "equator" at 
$x=L$.  The geometry of the system is sketched in Figure \ref{f1}.

For the form of $\eta$-quenching, we assume

\begin{equation}
  \eta=\frac{\eta_0}{1+\left(B/B_s\right)^2}\;,
  \label{eta_q}
\end{equation}
in which $\eta_0$ is the unquenched value, $B=B(x,z)$ is the toroidal field 
to be solved for, and $B_s$ is a constant, the value at which we assume 
$\eta$ has been reduced by 50\%.

We seek solutions for the toroidal field $B(x,z,t)$. The induction equation 
for this problem is given by 
\begin{eqnarray}
  \frac{\partial B}{\partial t}&=&B_x\frac{\partial u}{\partial x}+
  B_z\frac{\partial u}{\partial z}+\frac{\partial}{\partial x}\left(
  \frac{\eta_0}{1+\left(B/B_s\right)^2}\frac{\partial B}{\partial x}
  \right)\nonumber\\
  &+&\frac{\partial}{\partial z}\left(
  \frac{\eta_0}{1+\left(B/B_s\right)^2}\frac{\partial B}{\partial z}
  \right)\;,
  \label{induction1}
\end{eqnarray}
in which $u$ is the assumed differential rotation, and $B_x$, $B_z$ is the 
assumed poloidal field. We require
\begin{equation}
  \frac{\partial B_x}{\partial x}+\frac{\partial B_z}{\partial z}=0
  \label{divergence}
\end{equation}
but allow $B_x$ and/or $B_z$ to cross the "equatorial" and/or top boundaries. 
For simplicity we omit meridional circulation from this calculation.

There is a key transformation of the diffusion terms in equation 
(\ref{induction1}) that allows us to find analytical solutions, at least in 
the case of an assumed steady state. In particular from Burington's tables, 
page 38,
\begin{equation}
  \frac{1}{1+f^2}\frac{\mbox{d}f}{\mbox{d}x}=\frac{\mbox{d}}{\mbox{d}x}
  \tan^{-1}f
\end{equation}
from which we can write
\begin{eqnarray}
  \frac{\partial B}{\partial t}&=&B_x\frac{\partial u}{\partial x}+
  B_z\frac{\partial u}{\partial z}\nonumber\\
  &+&\eta_0 B_s\left(
  \frac{\partial^2}{\partial x^2}+\frac{\partial^2}{\partial z^2}\right)
  \tan^{-1}\left(\frac{B}{B_s}\right)\;.
  \label{induction2}
\end{eqnarray}
As a starting point, we seek steady solutions to equation (\ref{induction2}). 
If we define $a=\tan^{-1}(B/B_s)$ then equation (\ref{induction2}) reduces to
\begin{equation}
  \left(\frac{\partial^2}{\partial x^2}+\frac{\partial^2}{\partial z^2}\right)a
  =-\frac{1}{\eta_0}\left(\frac{B_x}{B_s}\frac{\partial u}{\partial x}+
  \frac{B_z}{B_s}\frac{\partial u}{\partial z}\right)\;,\label{ind_stat1}
\end{equation}
a Poisson-type equation for specific $B_x$, $B_z$ and $u$.

\subsection{Assumptions for differential rotation an poloidal field}
There are many choices for poloidal field and differential rotation. 
Trigonometric functions are particularly convenient and illustrative. 
We choose
\begin{eqnarray}
  u&=&-U_0\sin\left(\frac{3\pi x}{2L}\right)\sin\left(\frac{\pi z}{2H}
  \right)\;,\nonumber \\
  B_x&=&B_p\sin\left(\frac{\pi x}{2L}\right)\cos\left(\frac{\pi z}{2H}
  \right)\;,\nonumber \\
  B_z&=&-B_p\cos\left(\frac{\pi x}{2L}\right)\sin\left(\frac{\pi z}{2H}
  \right)\;, 
  \label{fields}
\end{eqnarray}
in which $U_0$ and $B_p$ are constant.  For these cases, $u$ has a maximum 
at $x=L$ (the "equator") and is zero at $x=2L/3$ ($30\degr$ "latitude") 
as well as at $x=0$ (the "pole"), 
very similar to the linear velocity of rotation on the Sun. There is also 
no differential rotation with $x$ at the bottom, and a maximum amount at 
the top (analogous to the solar tachocline). The poloidal field $B_z$ is a 
maximum at $x=0$, and zero at $x=L$, corresponding to a dipole field 
crossing the equator. $B_x$ is zero at $x=0$, and a maximum at $x=L$. By 
substitution, $B_x$, $B_z$ satisfy equation (\ref{divergence}). All these
fields are shown schematically in Figure \ref{f1}.
Obviously many other choices of poloidal field and differential rotation are
possible, but these seem particularly relevant to the solar case.

\begin{figure*}
  \resizebox{\hsize}{!}{\includegraphics{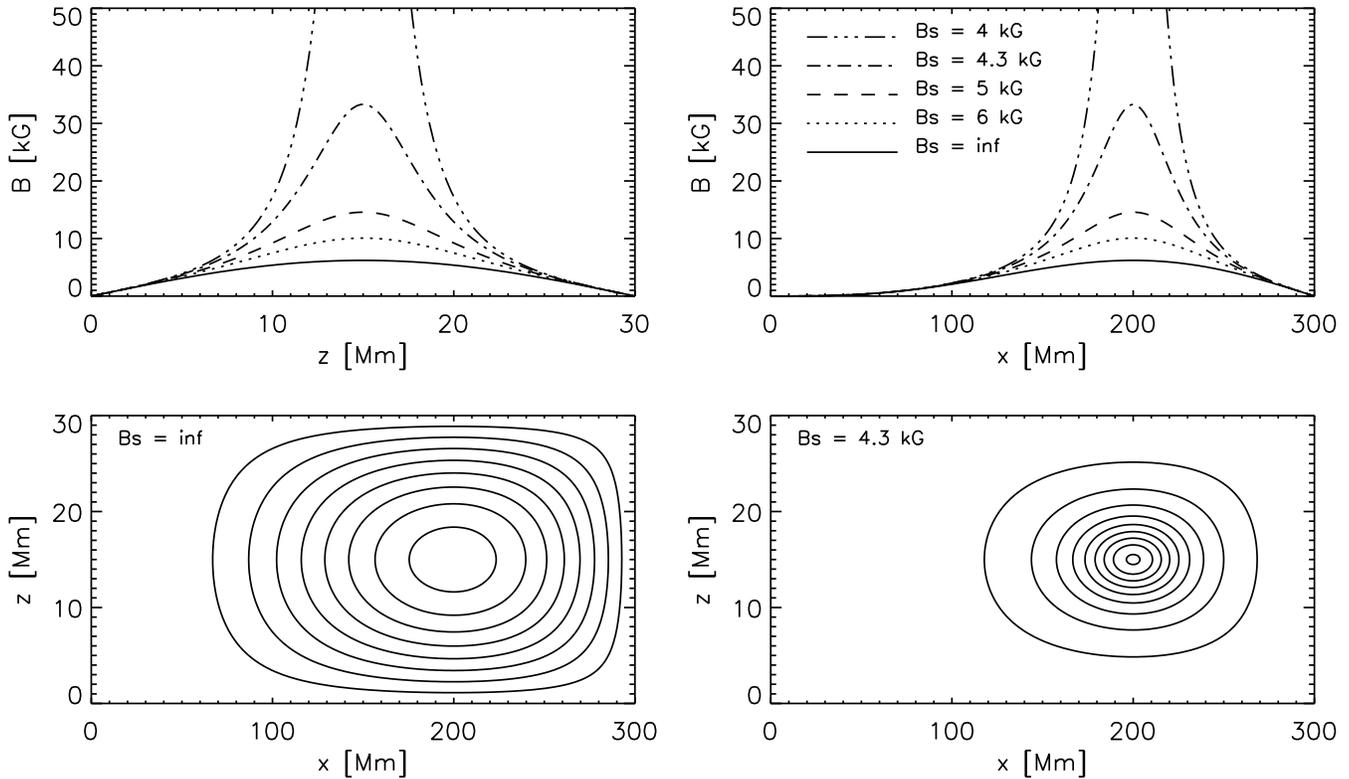}}
  \caption{Solutions of equation (\ref{sol_final}). The top panels show 
    a vertical cross section at $x=200\,\mbox{Mm}=2/3\,L$ (left) and a 
    horizontal
    cross section in the middle of the domain (right). The bottom panels show 
    contours of the toroidal
    field for two distinct cases. Common parameters are $R_m=3000$, $H/L=0.1$,
    and $B_p=1\,\mbox{kG}$. Solutions are shown for $B_s=\infty$, $6$, $5$, 
    $4.3$, and $4$ kG.
  }
  \label{f2}
\end{figure*}

\begin{figure*}
  \resizebox{\hsize}{!}{\includegraphics{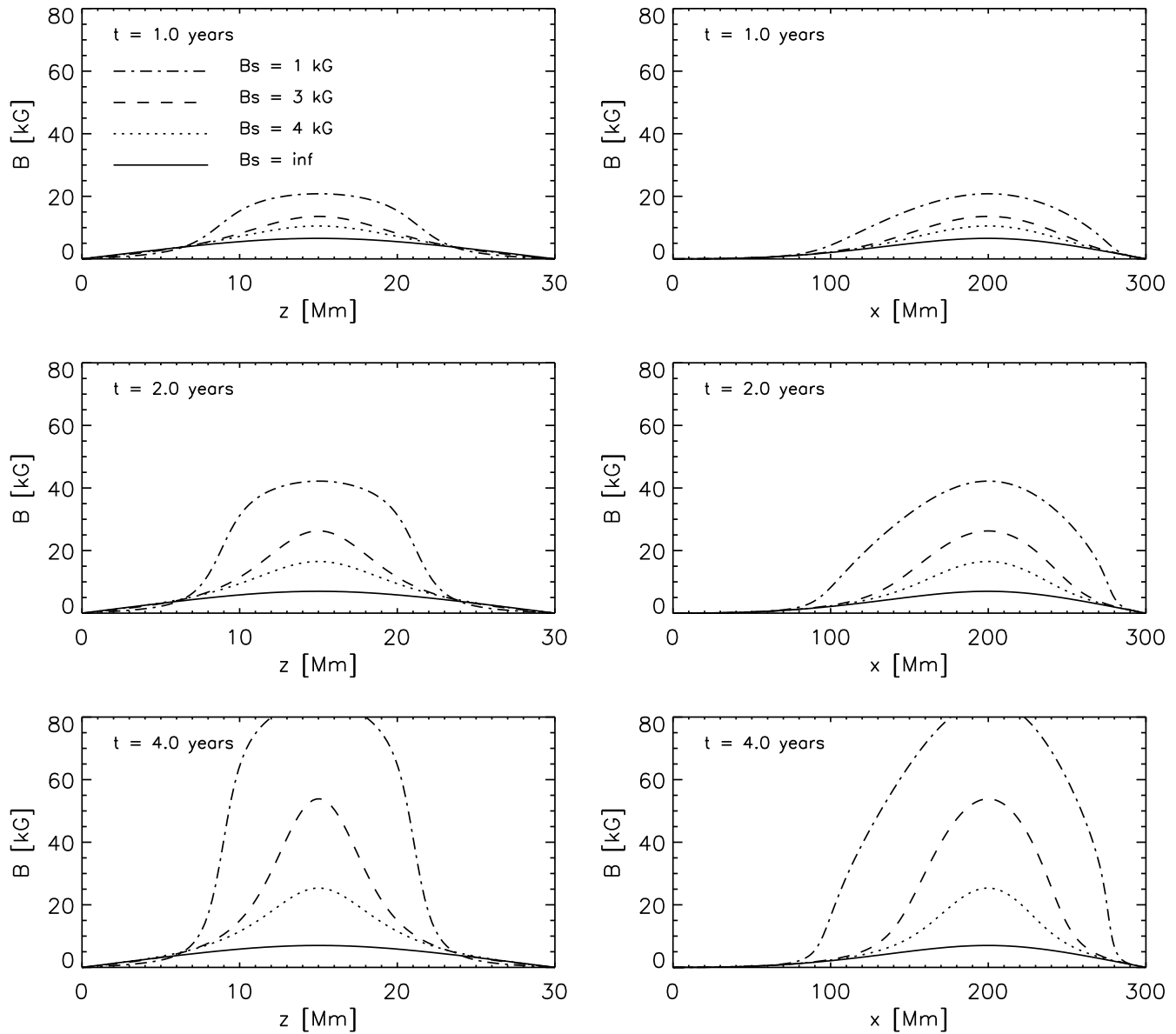}}
  \caption{Snapshots of time dependent solution after $1$, $2$, and $4$ 
    years. Shown are vertical cross sections at $x=200\,\mbox{Mm}=2/3\,L$ 
    (left panels) and horizontal cross sections (right panels) at 
    $z=15\,\mbox{Mm}=1/2\,H$ of the
    toroidal field. Each panel shows a solution with no $\eta$-quenching
    (solid), $\eta$-quenching with $B_s=4\,\mbox{kG}$ (dotted), 
    $\eta$-quenching with $B_s=3\,\mbox{kG}$ (dashed), and 
    $\eta$-quenching with $B_s=1\,\mbox{kG}$ (dashed-dotted).   
  }
  \label{f3}
\end{figure*}

\begin{figure*}
  \resizebox{\hsize}{!}{\includegraphics{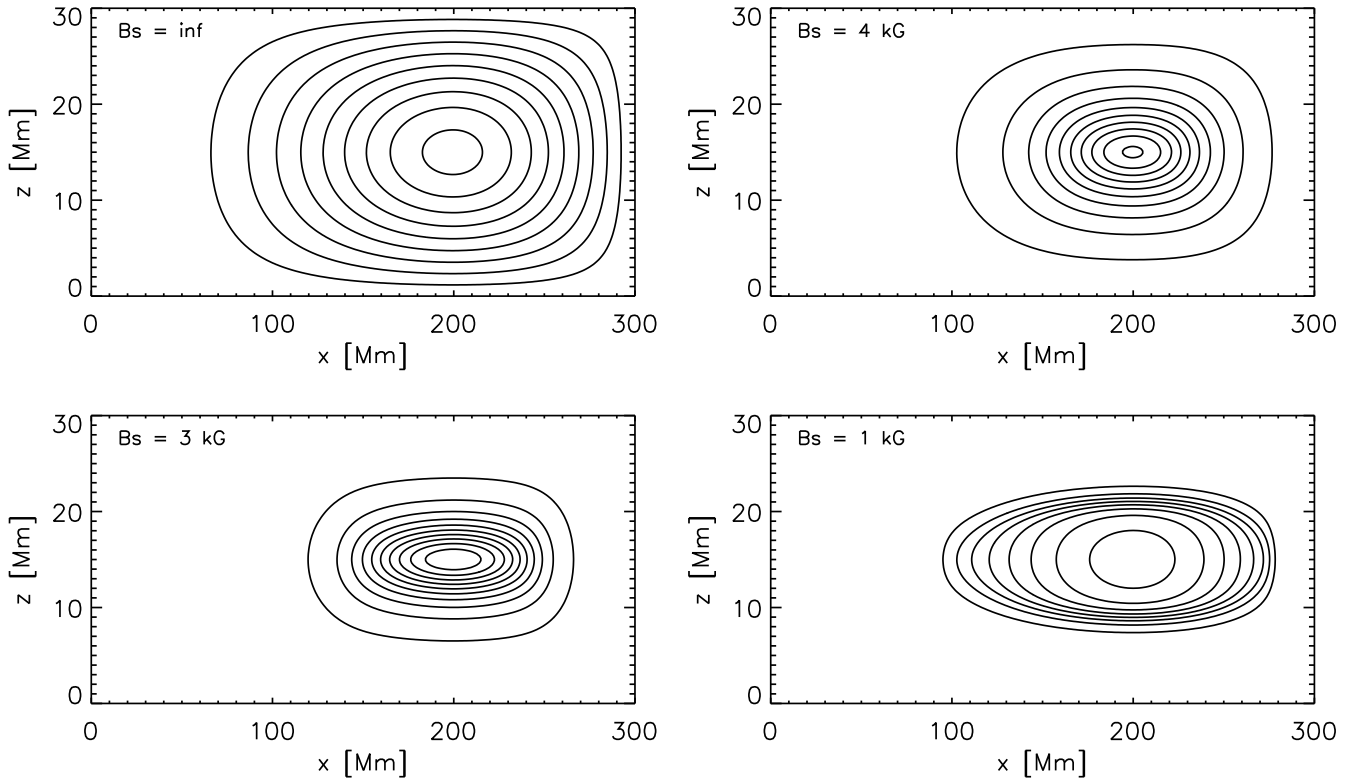}}
  \caption{Contours of the toroidal field after $4$ years of amplification
    for the different values of $B_s$ used for Figure \ref{f3}. 
    $\eta$-quenching tends to concentrate field close to the peak value.
  }
  \label{f4}
\end{figure*}

\subsection{Solution}
Substitutions of equations (\ref{fields}) into the right-hand side of 
(\ref{ind_stat1}) yields
\begin{eqnarray}
  \left(\frac{\partial^2}{\partial x^2}\right.&+&\left.\frac{\partial^2}
  {\partial z^2}\right)a=\frac{B_p}{B_s}\frac{\pi}{4L^2}R_m\sin
  \left(\frac{\pi z}{H}\right)\left[3\sin\left(\frac{\pi x}{2L}\right)\right.
  \nonumber\\
  &&\left.\cos\left(\frac{3\pi x}{2L}\right)
    -\cos\left(\frac{\pi x}{2L}\right)\sin\left(\frac{3\pi x}{2L}\right)\right]
  \label{ind_stat2}
\end{eqnarray}
in which $R_m=U_0\,L/\eta_0$ is a magnetic Reynolds number. By inspection, 
equation (\ref{ind_stat2}) should yield separable solutions
\begin{equation}
  a=g(x)\sin\left(\frac{\pi z}{H}\right)\;,\label{separation}
\end{equation}
from which we get 
\begin{equation}
  B=B_s\tan\left[g(x)\sin\left(\frac{\pi z}{H}\right)\right]\;.
  \label{sol_general}
\end{equation}
The tangent function is particularly critical in determining the form of the 
toroidal field $B$. The substitution of equation (\ref{separation}) into 
equation (\ref{ind_stat2}), and changing variables by letting 
$x=2L\lambda/\pi$ leads to
\begin{eqnarray}
\frac{\mbox{d}^2 g}{\mbox{d}\lambda^2}-4\left(\frac{L}{H}\right)^2 g&=&
\frac{B_p}{B_s}\frac{R_m}{\pi}\left[3\sin\lambda\cos 3\lambda\right.
  \nonumber\\&-&\left.\cos\lambda\sin 3\lambda\right]\label{dgl}
\end{eqnarray}
For the solar tachocline, we can argue that $L/H\gg 1$ (typical values are 
around $10-30$) to a first approximation, so we can ignore the second 
derivative term in equation (\ref{dgl}) (unless we need to include a boundary 
layer to satisfy a boundary condition). In that case
\begin{eqnarray}
  g(x)&=&\frac{B_p}{B_s}\left(\frac{H}{L}\right)^2\frac{R_m}{4\pi}\left[
    \cos\left(\frac{\pi x}{2L}\right)\sin\left(\frac{3\pi x}{2L}\right)\right.
    \nonumber\\
    &-&\left.3\sin\left(\frac{\pi x}{2L}\right)\cos\left(\frac{3\pi x}{2L}
    \right)\right]
\end{eqnarray}
and therefore the full solution for $B$ is 
\begin{eqnarray}
  B&=&B_s\tan\left\{\frac{B_p}{B_s}\left(\frac{H}{L}\right)^2\frac{R_m}{4\pi}
    \sin\left(\frac{\pi z}{H}\right)\left[
    \cos\left(\frac{\pi x}{2L}\right)\right.\right.\nonumber\\&&
    \left.\left.\sin\left(\frac{3\pi x}{2L}\right)
    -3\sin\left(\frac{\pi x}{2L}\right)\cos\left(\frac{3\pi x}{2L}
    \right)\right]\right\}\label{sol_final}
\end{eqnarray}
Thus we have found exact analytic solutions to a nonlinear partial 
differential equation!

\subsection{Qualitative Interpretation}
\label{qualitative}
The solution given in equation (\ref{sol_final}), a particular case of 
the more general solution given in equation (\ref{sol_general}), reveals 
many qualitative features, evident even without plotting. Of primary 
interest is that the induced toroidal field is proportional to the tangent 
of a function, of latitude and height, the magnetic Reynolds number, the 
ratio of assumed poloidal field to the quenching toroidal field, and the 
'aspect ratio' of the domain, which is small in the actual solar tachocline. 
As is well-known, when the argument $p$ of a tangent function is small,
then $\tan p\approx p$. In this limit, equation 
(\ref{sol_final}) becomes independent of the $\eta$-quenching field $B_s$. 
But $\tan p$ approaches infinity as the argument $p$ approaches $\pi/2$.
So we 
have the possibility of very large amplification of the toroidal field from 
an initially relatively small poloidal field, as well as compared to $B_s$. 

From equation (\ref{sol_final}) the induced toroidal field always vanishes 
at the 'pole' at $x=0$, as well as the 'equator ' at $x=L$, and the top and 
the bottom of the layer. This is true even though the poloidal field crosses 
both the top boundary at $z=H$ and the side boundary at $x=L$. Therefore the 
toroidal field is contained within the domain of induction, and peaks in 
amplitude at mid height, $z=H/2$. In addition, by inspection of equation 
(\ref{sol_final}), we can deduce that the argument of the tangent always has 
a peak at $x=2L/3$ the equivalent of $30\degr$ latitude on the Sun. 
Obviously this 
location is a function of our assumptions, but it follows from a differential 
rotation similar to that of the Sun, and a 'dipole-like' poloidal field, so 
its location near the latitude where spots are first seen in a new sunspot 
cycle may not be coincidence. $B_s$ is always positive, so for positive 
poloidal field $B_p$, the induced toroidal field is positive everywhere in 
the domain, even though not all terms in the induction forcing function 
(the right hand side) of equation (\ref{ind_stat1}) contribute with the same 
sign.

Given the above, then, we can expect larger amplification of the toroidal 
field in the interior of the domain, peaking at mid-depth and $x=2L/3$ or 
about $30\degr$ latitude. There can easily be a finite thickness and finite 
$x$ or latitude range where the argument of the tan exceeds $\pi/2$. 
What does this mean? In the section \ref{timedep}, we discuss reaching 
the steady solutions 
represented by equation (\ref{sol_final}) via time-dependent simulations 
that start from a state with zero toroidal field. We argue here that such 
solutions on the 'other side' of $\pi/2$ are not attainable in a finite time; 
the domain of such 'unreachable' solutions would be defined by the contour 
in the interior of our domain on which, for given $B_p$, $B_s$ and $R_m$, 
the argument of the $\tan=\pi/2$. On this contour, in effect steady solutions 
do not exist, because they would take an infinite time to reach the infinite 
field value. 

But so long as the argument of the tan remains smaller than $\pi/2$ in the 
whole domain, steady solutions are well defined throughout. Even in this case, 
however, it is of interest to know as a function of various parameters how 
long it takes a time-dependent solution to 'spin up' to the steady state. 
Obviously solutions that take the order of a sunspot cycle or longer to 
reach the steady state will generally not be realized in the sun, since the 
cycle progresses and the peak toroidal field moves in latitude.

The differences in predicted toroidal field amplitude as functions of $B_s$, 
$B_p$, and $R_m$ in equation (\ref{sol_final}) are straightforward to 
understand. Since the larger is $R_m$, the lower is the unquenched 
diffusivity, so we should expect larger toroidal fields to result from a 
given $B_p$. And obviously the larger is $B_p$ initially, the larger will be 
the resulting toroidal field. But the toroidal field is larger the smaller 
is $B_s$, since $B_s$ is in the denominator inside the tangent function and
in the numerator outside. This is because for smaller $B_s$, the local 
diffusivity  is reduced starting at lower toroidal field, and so locally the 
toroidal field is kept from diffusing away; hence it amplifies further. For 
low enough $B_s$, so strong is the induction that no steady state can be 
reached.

\subsection{Quantitative results}
Figure \ref{f2} shows solutions of equation (\ref{sol_final}) for the
common parameters $R_m=3000$, $H/L=0.1$, and $B_p=1\,\mbox{kG}$. To illustrate 
the amplification and concentration effect of $\eta$-quenching we have
shown solutions for $B_s=\infty$, $6$, $5$, $4.3$, and $4$ kG.
The top panels show a vertical cross section at $x=200\,\mbox{Mm}$ (left) and 
a horizontal cross section in the middle of the domain (right). 
The bottom panels show contours of the toroidal field for the cases with
$B_s=\infty$ (left) and $B_s=4.3\,\mbox{kG}$. The case with 
$B_s=4\,\mbox{kG}$ shown in the top panels is close to the transition toward
unbounded solutions (the peak field strength is around $200$ kG). 
Figure \ref{f2}
shows clearly the significant amplification and concentration effect
discussed qualitatively in section \ref{qualitative} before.

\subsection{Solar effects?} 
For plausible values of all the parameters in equation (\ref{sol_final}), 
will the amplification effect we have found be significant for the Sun, in 
particular the solar tachocline?  In the tachocline, we estimate that
$30\,\mbox{m}\,\mbox{s}^{-1}\leq U_0 \leq 100\,\mbox{m}\,\mbox{s}^{-1}$;
$10^9\,\mbox{cm}^2\,\mbox{s}^{-1}\leq \eta_0 \leq 10^{11}\,\mbox{cm}^2\,
\mbox{s}^{-1}$, from weak overshooting turbulence there; 
$3\times 10^5\,\mbox{km}\leq L\leq 10^6\,\mbox{km}$. These values lead to
$10^3\leq R_m\leq 10^6$. For a tachocline of thickness  
$H\sim 3\times 10^4\,\mbox{km}$, $0.03\leq H/L\leq 0.1$. 
Plausible ranges for $B_p$ and $B_s$ are $5\times 10^2\,\mbox{G}\leq B_p
\leq 5\times 10^3\,\mbox{G}$ and $10^3\,\mbox{G}\leq B_s\leq 10^4\,\mbox{G}$.
Then
\begin{equation}
4\times 10^{-3}\leq\frac{1}{4\pi}\frac{B_p}{B_s}\left(\frac{H}{L}\right)^2 
R_m\leq 4\times 10^3\label{paramrange}
\end{equation}

so in the Sun there is a full range of amplification possible, from extremely 
small to extremely large, depending on the values of the various parameters 
we choose. Given that the peak value of the trig functions inside of equation 
(\ref{sol_final}) is about 2.6, even the smallest argument of the tangent is  
$\sim 10^{-2}$, with $\pi/2=1.57$  corresponding to infinite amplification; 
mid-range values for all parameters in (\ref{paramrange}) leads to an 
argument of the tangent for the solar tachocline of $\sim 10$, a factor of 
$6$ above the first infinity. Thus it is easy to have solar conditions for 
which the amplification is large.

\section{Time dependent solutions}
\label{timedep}
\begin{figure*}
  \resizebox{\hsize}{!}{\includegraphics{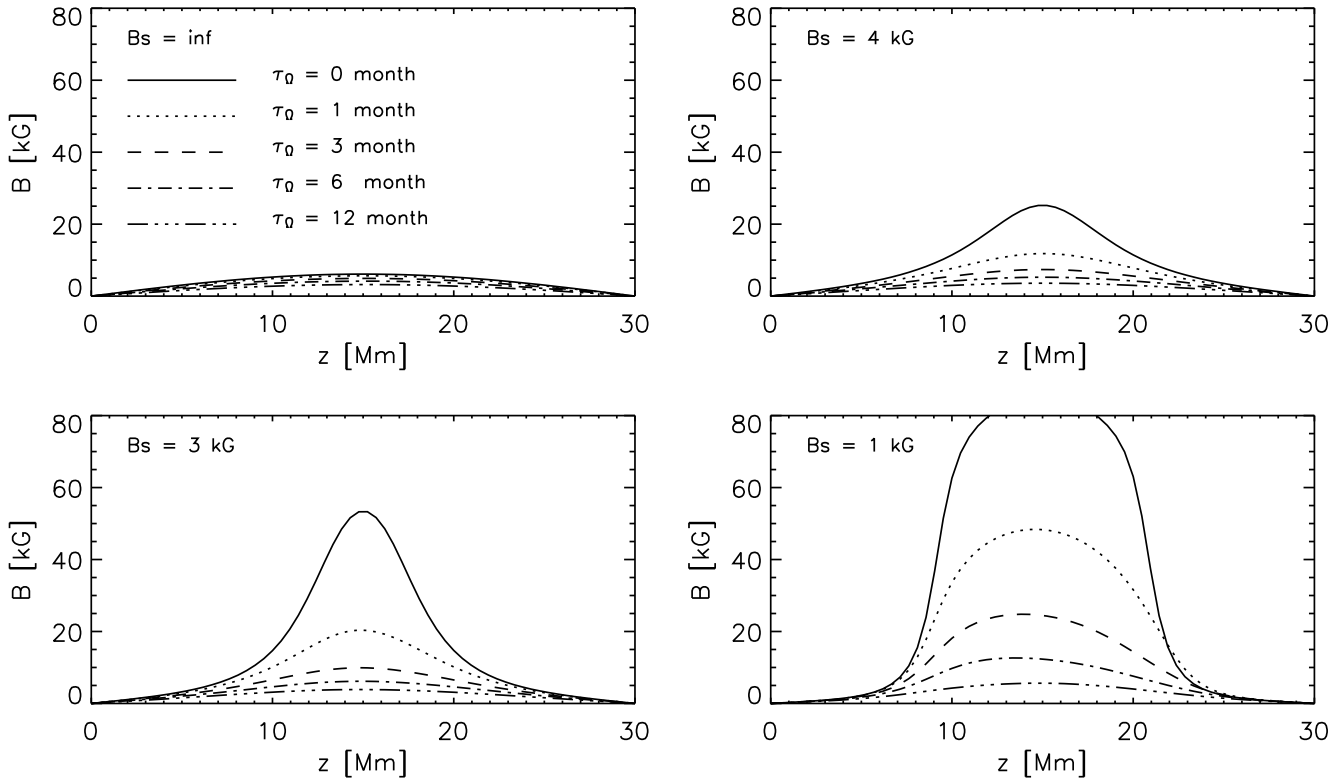}}
  \caption{Influence of feedback on differential rotation on amplification
    process. We show snapshots of the solutions after $4$ years. In each
    panel the solid line indicates the solutions presented already in Figure 
    \ref{f3}, the dotted, dashed, dashed-dotted, and triple-dashed-dotted
    lines represent solutions including feedback through Lorentz force and
    a restoration time scale for the differential rotation of $\tau=1$, $3$, 
    $6$, $12\,\mbox{months}$, respectively.
  }
  \label{f5}
\end{figure*}

\begin{figure*}
  \resizebox{\hsize}{!}{\includegraphics{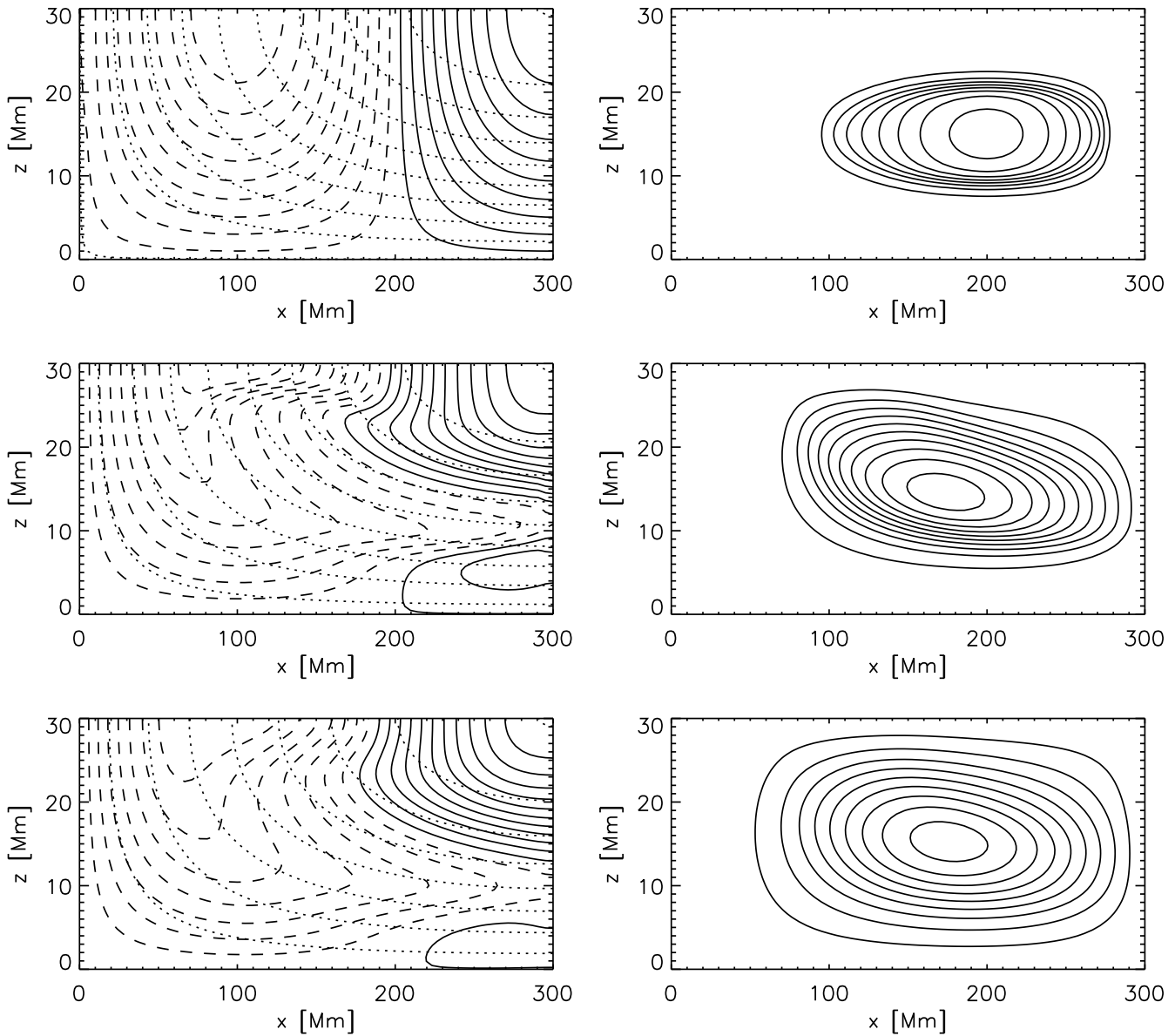}}
  \caption{Contours of differential rotation (left panels) and toroidal field
    (right panels). The dotted curves in the left panels are the poloidal field
    lines. The top panels show the reference solution with no
    feedback, the middle panels the solution with $\tau=3\,\mbox{months}$,
    and the bottom panels the solution with $\tau=12\,\mbox{months}$.
    All snapshots are after $4$ years of amplification time.
  }
  \label{f6}
\end{figure*}
There are at least two effects not included in the solutions in 
(\ref{sol_final}) that could significantly diminish the amplification effect 
on time scales relevant to a sunspot cycle. One, that we have already 
mentioned, is that the steady state is not reached before the solar cycle 
toroidal field moves to a lower latitude. The second is that as the toroidal 
field grows, there should develop a {\boldmath$j\times B$} reaction force 
which damps the differential rotation responsible for the induction. Here we 
explore both of these effects with time dependent numerical calculations.

We first focus on time dependent solutions only solving the induction
equation and neglecting the feedback of the magnetic field on the shear
flow. To this end we solve equation (5) using a finite difference code. 
We tested our numerical setup by comparing the stationary solution obtained
through our code with the solution shown in equation (13). For this comparison
we have chosen the parameters in our numerical setup such that the argument of 
the tangent in equation (\ref{sol_final}) does not exceed $\pi/2$ anywhere in 
the domain to ensure the existence of a stationary solution.

Common parameters for the time dependent solutions discussed below are:
$L=3\times 10^5\,\mbox{km}$, $H=3\times 10^4\,\mbox{km}$, 
$\eta_0=10^{11}\,\mbox{cm}^2\,\mbox{s}^{-1}$ , 
$U_0=100\,\mbox{m}\,\mbox{s}^{-1}$, and $B_p=1\,\mbox{kG}$.
With this choice of parameters we have $R_m=3000$, $H/L=0.1$ and from 
equation (13) it can be expected that the argument of the tangent exceeds
$\pi/2$ if $B_s=4\,\mbox{kG}$. Figure \ref{f3} shows snapshots of time 
dependent solutions after $1$ year, $2$ years, and $4$ years. 
The panels on the left show vertical cross sections at $x=200\,\mbox{Mm}$, 
where the field peaks in the horizontal coordinate. 
The panels on the right show 
horizontal cross sections in the middle of the domain ($z=15\,\mbox{Mm}$). 
We show a solutions with no $\eta$-quenching (solid line)
and solutions with a value of $B_s$ of $4$ kG (dotted), $3$ kG (dashed), and 
$1$ kG (dashed-dotted). We can achieve similar toroidal field amplification
patterns with $B_s$ set much higher simply by assuming a smaller $\eta_0$
and therefore a larger $R_m$. As explained above the solution with 
$B_s=4\,\mbox{kG}$ marks the transition from asymptotically stationary to 
unbounded solutions. The solution with no $\eta$-quenching reaches an 
asymptotic field strength of around $6$ kG, whereas the solution with 
$B_s=1\,\mbox{kG}$ reaches around $80$ kG after $4$ years
of amplification through the shear flow. Even the solution with moderate
quenching $B_s=4\,\mbox{kG}$ reaches after $4$ years a value more than three 
times larger than the reference solution with no $\eta$-quenching. Therefore
it can be expected that the influence of $\eta$-quenching becomes visible 
even on time scales of the solar cycle. 

As an additional effect, the 
$\eta$-quenching leads to a confinement of the magnetic field around the peak
value, leading to a magnetic layer of less extent in radius and latitude 
compared to the reference solution with no quenching. The effect of 
confinement is largest for values of $B_s$ close to the critical value for
which only the most central part of the magnetic field gets significantly
amplified ($B_s=3\,\mbox{kG}$ and $4\,\mbox{kG}$), whereas strong quenching 
($B_s=1\,\mbox{kG}$) leads to a significant amplification of a much broader 
field profile.

Figure \ref{f4} shows contour plots of the toroidal field 
after $4$ years for the cases shown in Figure \ref{f3}. The spacing of the 
contours is equidistant, ranging from $0$ to the maximum value of each 
particular case. The greater concentration of toroidal flux around the center
of the layer is clearly evident for $B_s=3$ and $4$ kG.

The time dependent solutions discussed so far show clearly that a significant
amplification of a $1$ kG poloidal field to several tens of kG toroidal field 
can be achieved within a few years, if $\eta$-quenching is taken into 
consideration.
However, these solutions assume that the feedback on the differential rotation
through the Lorentz force can be neglected. In order to address this feedback
in time dependent simulations we solve together with the induction equation
an equation for the differential rotation of the form
\begin{equation}
\frac{\partial u}{\partial t}=-\frac{u-u_0}{\tau}+\frac{1}{4\pi\varrho}
\left(B_x\frac{\partial B}{\partial x}+B_z\frac{\partial B}{\partial z}\right)
\label{omega}
\end{equation}
In equation (\ref{omega}) we include a drag-type forcing term which replenishes
the differential rotation from the convection zone above on a time scale 
$\tau$. $u_0$ is the differential rotation in the case of no Lorentz force 
feedback. For the density $\varrho$ we use a value of 
$210\,\mbox{kg}\,\mbox{m}^{-3}$ to represent the overshoot region. 
Since there are currently
no theories for the differential rotation in the tachocline, which would
help to determine a reasonable value for $\tau$, we treat $\tau$ as a free 
parameter
and study the dependence of the solution on particular choices of values
in the range of $1$ month to $1$ year. Differential rotation models for the
entire convection zone \citep{Miesch:etal:2000,Brun:Toomre:2002,Rempel:2005} 
typically predict a time scale of several years; however, this time scale 
should be shorter if only the tachocline is considered.

Figure \ref{f5} shows the influence of the feedback on differential rotation
for different values of $\tau$ ranging from $1$ month to $1$ year. In each 
panel is shown the vertical profile of the toroidal field at 
$x=200\,\mbox{Mm}$; 
the solid line corresponds to the solutions shown in Figure \ref{f3} after $4$ 
years. Even if a very short time scale for $\tau$ of $1$ month is used, 
the peak field strength reached is reduced by a factor of $2$ in the cases 
with quenching of $\eta$ (the case with no $\eta$-quenching shows a much 
weaker dependence on $\tau$ since the field 
strength is very weak anyway for this choice of parameters). For the choice of 
$\tau=1\,\mbox{year}$ the peak field strength barely exceeds $5$ kG. However, 
the tendency that $\eta$-quenching leads to stronger and more confined field 
is still visible in all cases shown.

Figure \ref{f6} shows the profile of differential rotation (left panels) and 
toroidal magnetic field (right panels) for the cases with $B_s=1\,\mbox{kG}$. 
The top panels show the reference solution with no feedback, the middle panels 
the solution with $\tau=3\,\mbox{month}$ and the bottom panels the solution 
with $\tau=1\,\mbox{year}$. In the left panels solid lines indicate positive 
values of $u$, dashed lines
negative values. We overplotted the field lines of the poloidal field
to show how the contours of the velocity field get aligned with the poloidal
field lines to minimize the induction effect (Ferraros's law of iso-rotation).
As a consequence, the feedback through the Lorentz force moves the shear layer
of the differential rotation upward close to the equator ($x=300\,\mbox{Mm}$) 
and downward close to the pole ($x=0\,\mbox{Mm}$), leading to a significant 
reduction of
the latitudinal shear in the middle of the domain, where the toroidal field
peaks. 

To summarize, $\eta$-quenching has a significant influence on the field 
amplification, even if the time available for amplification is limited to
a few years to reflect the solar cycle variability. Including the feedback
on differential rotation through Lorentz force leads to a significant decrease
in the peak field strength reached; however, the influence of $\eta$-quenching
remains visible in most cases. Unless time scales of less than $1$ month for 
the
replenishment of differential in the tachocline are assumed it seems impossible
to amplify a magnetic field to $100\,\mbox{kG}$ as inferred by simulations of 
rising 
flux tubes. Other amplification mechanisms that do not rely on differential
rotation as energy source, as for example the explosion of magnetic flux tubes 
\citep{Rempel:Schuessler:2001} or the downward draining of their interiors, 
most probably also play an important role.

\section{Conclusions and discussion}
We have demonstrated with a simple Cartesian model that the damping of 
turbulent 
magnetic diffusivity by the growth of strong toroidal fields is a powerful 
mechanism for making such fields even stronger and more concentrated. Even 
though the expected reaction of these stronger fields on the differential 
rotation that induced them can limit the amplitude of this effect, it should 
remain an important contributor to the overall operation of the dynamo inside 
the Sun.

Our result is qualitatively consistent with the effect common to all flux
transport and interface dynamos for the sun, in which the high toroidal
field in the tachocline is achieved because of the low magnetic diffusivity
there. In that case, the low diffusivity comes from the assumed weak 
turbulence there, while in the present work the toroidal field is assumed
to locally reduce the turbulence. 
This quenching has already been shown by \citet{Tobias:1996} to have profound 
effects on, for example, dynamo mode type, when applied to more idealized 
interface dynamos considered for the bottom of the convection zone.

The role this toroidal field amplification mechanism plays in current dynamo 
models for the Sun should be tested. While our model used Cartesian geometry 
because of its simplicity and because it yielded useful analytical 
solutions, we 
have done sample calculations in spherical geometry that show similar effects 
will be present there. 

It will be interesting to see even in the kinematic dynamo regime how much 
concentration and amplification of toroidal field occurs when a real dynamo 
solution is advancing through a simulated solar cycle, so the toroidal field 
is continually moving toward the equator. This effect could also be simulated 
with 
our simple Cartesian model by taking poloidal fields that represent various 
phases of a solar cycle; the case we have chosen corresponds roughly to the 
maximum phase, since the poloidal field peaks at the pole and there are no 
separate 'old' and 'new' cycle poloidal fields present. It is clear from our 
basic equations that analytic solutions should also exist for these other 
cycle phases. An important factor limiting toroidal field amplification will 
be simply 
how long in reality it takes for the imposed poloidal field to move in $x$ 
(latitude) enough to change significantly the location where induction of 
toroidal field is strongest.

\acknowledgements
The authors thank Mausumi Dikpati for helpful comments on a draft of
this paper. We also thank the anonymous referee for a very helpful 
review.

\bibliographystyle{natbib/apj}
\bibliography{natbib/apj-jour,natbib/papref}

\end{document}